\documentclass[11pt,twoside]{article}
\usepackage{asp2004}
\usepackage{psfig}
\usepackage{epsf}
\usepackage{graphics}
\usepackage{lscape}

\newcommand{\cmjj}{\mbox{${\rm cm^{-2}}$}}
\newcommand{\etal}{et al.}
\newcommand{\hI}{\mbox{H\,I}}

\newcommand{\lya}{\mbox{${\rm Ly}\alpha$}}

\def\edcomment#1{\iffalse\marginpar{\raggedright\sl#1\/}\else\relax\fi}
\reversemarginpar

\begin{document}
\title{Extended Neutral Gas Around \boldmath $z\sim 0.5$ Galaxies: Properties 
of Damped \boldmath \lya\ Absorbing Galaxies}
\author{Hsiao-Wen Chen}
\affil{Center for Space Research, Massachusetts Institute of Technology, 
Cambridge, MA 02139, U.S.A.}

\begin{abstract}

I review current results from searching for galaxies giving rise to damped
\lya\ absorbers (DLAs) at $z<1$.  Using 14 confirmed DLA galaxies, I further 
show that intermediate-redshift galaxies possess large \hI\ envelope out to 
$24-30\ h^{-1}$ kpc radius.  The photometric and spectral properties of these 
galaxies confirm that DLA galaxies are drawn from the typical field population,
and not from a separate population of low surface brightness or dwarf galaxies.
Comparisons of the ISM abundances of the DLA galaxies and the metallicities of 
the absorbers at large galactic radii suggest that some DLAs originate in the 
relatively unevolved outskirts of galactic disks.

\end{abstract}
\thispagestyle{plain}

\section{Background}

  Damped \lya\ absorbers (DLAs) observed in the spectra of background QSOs 
probe neutral gas regions commonly seen in nearby galaxies (with \hI\ column 
density $N(\hI)\ge 2\times 10^{20}$ \cmjj).  In principle, these DLAs offer a 
means of studying the galaxy population at high redshift using galaxies 
selected uniformly based on known neutral gas content, rather than optical 
brightness or color.  Optical spectroscopic surveys for DLAs have demonstrated
that DLAs dominate the mass density of neutral gas in the universe and that 
they contain roughly enough gas at $z=3.5$ to form the bulk of the stars in 
present-day galaxies (Wolfe \etal\ 1995; Storrie-Lombardi \& Wolfe 2000; 
P\'eroux \etal\ 2002; Prochaska \& Herbert-Fort 2004).  But chemical abundance
analyses of DLAs at redshift $z < 1.6$ yield sub-solar metallicities in the 
neutral gaseous clouds (Pettini \etal\ 1999; Prochaska \etal\ 2003), suggesting
that they do not trace the bulk of star formation and are likely to represent a
biased sample of galaxies.  Because metallicity varies as functions of 
morphological type and galactocentric distance, it is impossible to understand
the origin of DLAs without first identifying the absorbing galaxies.

\begin{tiny}
\begin{table}[th]
\caption{Summary of Confirmed DLA Galaxies at $z<1$} 
\smallskip
\begin{center}
{\tiny 
\begin{tabular}{p{0.8 in}crcrrrcc}
\hline
\hline
 & & & & \multicolumn{1}{c}{$\Delta\theta$} &
\multicolumn{1}{c}{$\rho\times h^a$} & & $M_{AB}(B)^b$ & \\
\multicolumn{1}{c}{QSO} & $z_{\rm QSO}$ & $z_{\rm DLA}$ & $\log N(\hI)$ &
\multicolumn{1}{c}{($''$)} & \multicolumn{1}{c}{(kpc)} &
\multicolumn{1}{c}{$AB$} & $-5\,\log\,h$ & Morphology \\
\multicolumn{1}{c}{(1)} & (2) & (3) & (4) & \multicolumn{1}{c}{(5)} &
\multicolumn{1}{c}{(6)} & \multicolumn{1}{c}{(7)} & (8) & (9) \\
\hline
TON 1480 \dotfill & 0.614 & 0.0036 & 20.34 & 114.0 & 5.94 & $B=11.5$ & $-18.7$ & S0 \\
HS1543$+$5921 \dotfill & 0.807 & 0.009 & 20.35 & 2.4 & 0.31 & $R=16.5$ & $-15.3$ & LSB \\
PKS0439$-$433 \dotfill & 0.593 & 0.101 & 19.85 & 3.9 & 5.13 & $I=17.2$ & $-19.6$ & disk \\
Q0738$+$313 \dotfill & 0.635 & 0.2212 & 20.90 & 5.7 & 14.23 & $I=20.9$ & $-17.7$ & compact \\
Q0809$+$483 \dotfill & 0.871 & 0.4368 & 20.80 & 1.5 & 5.9 & $R=19.9$ & $-20.3$ & disk \\
B2\,0827$+$243 \dotfill & 0.939 & 0.525 & 20.30 & 5.8 & 25.42 & $R=21.0$ & $-20.0$ & disk \\
PKS1629$+$120 \dotfill & 1.795 & 0.532 & 20.70 & 3.0 & 13.24 & $R=21.6$ & $-19.2$ & disk \\
LBQS0058$+$0155 \dotfill & 1.954 & 0.613 & 20.08 & 1.2 & 5.67 & $R=23.7$ & $-17.6$ & disk \\
HE1122$-$1649 \dotfill & 2.400 & 0.681 & 20.45 & 3.6 & 17.66 & $I=22.4$ & $-18.8$ & compact \\
FBQS\,0051$+$0041 \dotfill & 1.190 & 0.740 & 20.40 & 3.3 & 16.87 & $I=22.4$ & $-18.6$ & compact \\
EX0302$-$2223 \dotfill & 1.400 & 1.001 & 20.36 & 3.3 & 18.65 & $R=23.2$ & $-19.3$ & Irr \\
\hline 
PKS 1127$-$145\dotfill & 1.187 & 0.313 & 21.71 &  3.8 & 12.2 & $R=22.4$ & $-16.8$ & Irr \\
                       &       &       &       &  3.8 & 12.2 &     22.1 & $-17.4$ & compact \\
                       &       &       &       &  9.8 & 31.5 &     19.1 & $-20.1$ & disk \\
                       &       &       &       & 17.5 & 56.2 &     18.9 & $-20.3$ & disk \\
AO0235$+$164 \dotfill & 0.940 & 0.524 & 21.70 & 2.1 &  9.4 & $I=20.2$ & $-20.3$ & compact \\
                      &       &       &       & 6.4 & 28.0 & $I=20.9$ & $-19.7$ & compact \\
FBQS\,1137$+$3907 \dotfill & 1.020 & 0.719 & 21.10 & 2.5 & 12.6 & $K=21.4$ & $-18.8$ & compact \\
                           &       &       &       & 1.5 &  7.6 & $K=21.7$ & $-18.5$ & Irr \\
\hline
\multicolumn{9}{l}{a.\ A $\Lambda$ cosmology, $\Omega_{\rm M}=0.3$ and 
$\Omega_\Lambda = 0.7$ with $h = H_0/(100 \ {\rm km} \ {\rm s}^{-1}\ {\rm 
Mpc}^{-1})$ is adopted throughout.} \\
\multicolumn{9}{l}{b.\ $M_{{\rm AB}_*}(B)-5\log h = -19.6$ (Ellis \etal\ 
1996).}
\end{tabular}}
\end{center}
\end{table}
\end{tiny}

\begin{tiny}
\begin{table}[!ht]
\caption{Summary of Candidate DLA Galaxies at $z<1$}
\smallskip
\begin{center}
{\tiny
\begin{tabular}{p{0.8 in}crcrrrcc}
\hline
\hline
 & & & & \multicolumn{1}{c}{$\Delta\theta$} &
\multicolumn{1}{c}{$\rho\times h$} & & $M_{AB}(B)$ & \\
\multicolumn{1}{c}{QSO} & $z_{\rm QSO}$ & $z_{\rm DLA}$ & $\log N(\hI)$ &
\multicolumn{1}{c}{($''$)} & \multicolumn{1}{c}{(kpc)} &
\multicolumn{1}{c}{$AB$} & $-5\,\log\,h$ & References \\
\multicolumn{1}{c}{(1)} & (2) & (3) & (4) & \multicolumn{1}{c}{(5)} &
\multicolumn{1}{c}{(6)} & \multicolumn{1}{c}{(7)} & (8) & (9) \\
\hline
Q0738$+$313 \dotfill & 0.635 & 0.0912 & 21.18 & ... & ... & $K>17.8$ & $>-18.8$ & 1,2 \\
PKS0952$+$179\dotfill & 1.472 & 0.2390 & 21.32 & ... & ... & ... & ... & 3 \\
PKS1229$-$021\dotfill & 1.038 & 0.3950 & 20.60 & 1.4 & $\approx 5.2$ & $R=22.1$ & $\approx -18.4$ & 4 \\
Q1209$+$107\dotfill & 2.191 & 0.6295 & 20.48 & 1.6 & $\approx 7.7$ & $R=21.6$ & $\approx -20.5$ & 5 \\
PKS1622$+$23\dotfill & 0.927 & 0.6563 & 20.36 & ... & ... & $R>24.5$ & $>-16.9$ & 6 \\
Q1328$+$307\dotfill & 0.849 & 0.692 & 21.30 & 2.0 & $\approx 10$ & $I=22.1$ & $\approx -19.3$ & 5 \\
PKS0454$+$039\dotfill & 1.345 & 0.8596 & 20.76 & 0.8 & $\approx 4.3$ & $R=24.2$ & $\approx -19.0$ & 4,5 \\
\hline
\multicolumn{9}{l}{1.\ Turnshek \etal\ (2001); 2.\ Cohen (2001); 3.\ Rao 
\etal\ (2003); 4.\ Steidel \etal\ (1994); 5.\ Le Brun \etal\ (1997); }\\
\multicolumn{9}{l}{6.\ Steidel \etal\ (1997).} \\
\end{tabular}}
\end{center}
\end{table}
\end{tiny}

   Identifying the optical counterpart of the DLAs has been a challenging task,
because these galaxies are faint and located at small projected distances to 
the background, bright QSOs (as implied by the intrinsic high column density of
the absorbers).  Le Brun \etal\ (1997) reported candidate absorbing galaxies
toward seven QSO lines of sight with known DLAs based on high spatial 
resolution HST/WFPC2 images.  These galaxies exhibit a wide range in 
morphological types, from luminous spiral galaxies, to compact objects, and to 
low surface brightness galaxies.  In two fields, they also identify possible 
groups of galaxies that are associated with the DLAs.  In contrast, Rao \etal\ 
(2003) collected a sample of 14 candidate or confirmed DLA galaxies, and
argued that low-luminosity dwarf galaxies dominate the DLA galaxy population 
with $\approx 50$\% of the DLAs originating in galaxies of luminosity $L\le 
0.25\,L_*$.

   Despite extensive searches in the past decade, the number of confirmed DLA 
galaxies remains relatively small.  Of all the 23 DLA systems known at $z\le 
1$,  14 have been identified with their host galaxies using either 
spectroscopic or photometric redshift techniques (see section 2 for a complete
list).  Here I will first review current results from searching for $z<1$ DLA 
galaxies.  Then I will compare the optical properties of the absorbing 
galaxies, such as $B$-band luminosity, stellar content, and rotation velocity, 
with known properties of the absorbers, such as $N(\hI)$ and metallicity.

\section{A Sample of DLA Galaxies at $z<1$}

  A list of 14 DLA galaxies confirmed using either spectroscopic or photometric
redshift techniques is presented in Table 1 (see also Chen \& Lanzetta 2003 for
a list of references and Lacy \etal\ 2003 for new additions).  It is 
interesting to note that the three highest-$N(\hI)$ DLAs are found to be 
associated with a group of galaxies within a small projected distance (the last
three DLAs in Table 1).  In addition, the DLAs toward TON\,1480 and 
HS1543$+$5921 were discovered in targeted searches of DLAs, with a prior 
knowledge of the presence of a foreground galaxy within a small radius from the
sightline, while the rest of the absorbers were identified in a survey of 
random QSO sightlines.

  Additional efforts were made to search for the absorbing galaxies of seven 
DLAs listed in Table 2.  Candidate galaxies were found for the DLAs toward 
PKS0952$+$179, PKS1229$-$021, Q1209$+$107, Q1328$+$307, and PKS0454$+$039 based
on their close proximity to the QSO sightlines, while no candidates were found 
for the DLAs toward Q0738$+$313 and PKS1622$+$23 after extensive surveys.  

  The collection of 14 confirmed DLA galaxies represents the first large 
DLA-selected galaxy sample, which allows us (1) to determine the neutral gas 
cross-section of intermediate-redshift galaxies; (2) to establish an empirical 
correlation between the kinematics and metallicity of the absorbers and those 
of the absorbing galaxies; and (3) to study the disk population at intermediate
redshift using galaxies selected uniformly based on known neutral gas content, 
rather than optical brightness or color.

\section{\hI\ Extent of Intermediate-redshift Galaxies}

\begin{figure}
\plotfiddle{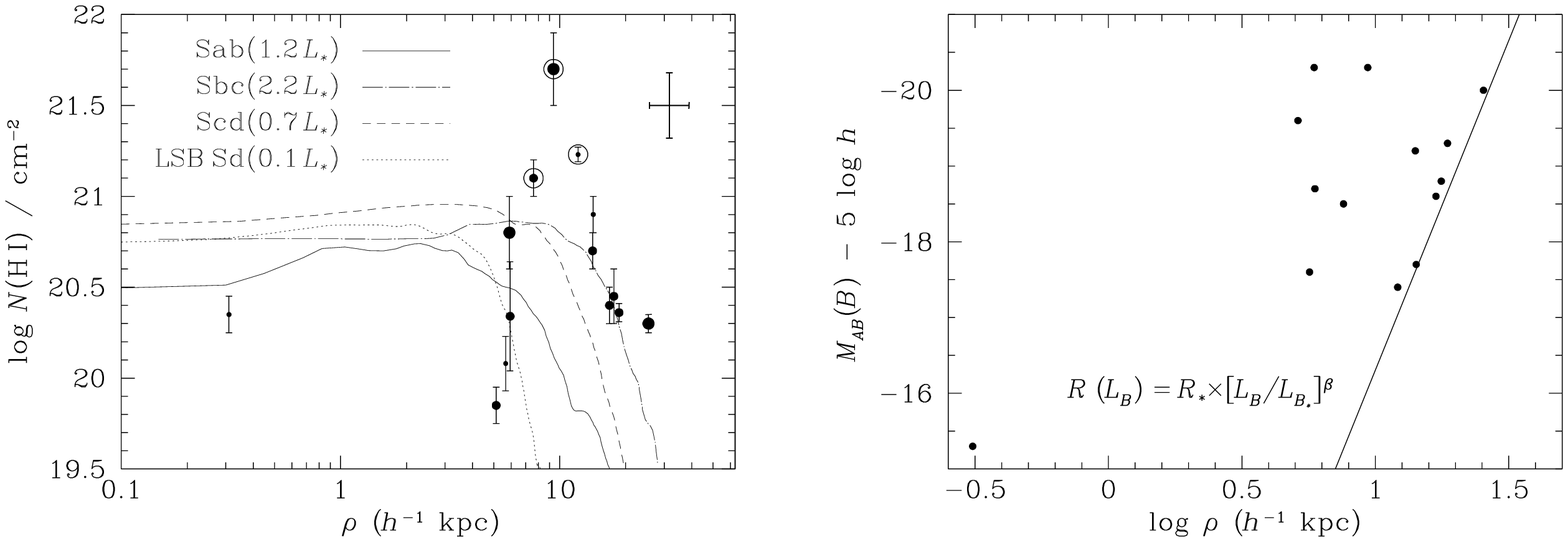}{1.7in}{0}{50}{50}{-190.0}{-10.0}
\caption[]{Left: $N(\hI)$ distribution versus galaxy impact parameter $\rho$ 
from 14 DLA galaxies (points), compared to the mean \hI\ profiles of nearby 
galaxies of different morphological type and intrinsic luminosity shown in 
curves.  Right: The distribution of $B$-band absolute magnitude versus $\rho$ 
for the 14 galaxies.  The solid line shows the best-fit scaling relation that 
describes the apparent envelope stretching to larger $\rho$ at brighter 
$M_{AB}(B)$.}
\end{figure}

  Figure 1 shows $N(\hI)$ versus $\rho$ for the 14 galaxy and DLA pairs in the
left panel.  Mean \hI\ surface density profiles measured from 21-cm 
observations of nearby galaxies of different morphological type and intrinsic 
luminosity are presented in different curves (the mean \hI\ profiles for Sab, 
Sbc, and Scd galaxies were digitized from Cayatte \etal\ 1994; the curve for 
LSB Sd was provided by Uson \& Matthews 2003).  The scatter of $N(\hI)$ at the 
Holmberg radii and the scatter of the neutral gaseous extent at $N(\hI) =
10^{20}$ \cmjj\ from 21-cm data for Sd-type galaxies are marked by the error 
bar in the upper-right corner (Cayatte \etal\ 1994).  The three DLAs found in 
groups of galaxies are marked in circles.  The size of the points indicates the
intrinsic brightness of the galaxies: $M_{AB}(B) - 5\log h \le -19.6$ (large), 
$-19.6 < M_{AB}(B) - 5\log h \le -18$ (medium), and $M_{AB}(B) - 5\log h > -18$
(small).  Despite the apparent large scatter in the $N(\hI)$ distribution, we 
see two interesting features.  First, the $N(\hI)$ in DLAs is not grossly 
different from the mean \hI\ distribution of nearby galaxies, although we note 
that measurements obtained in 21-cm observations are smoothed over a finite 
beam size.  Second, while the \hI\ extent of these intermediate-redshift 
galaxies appears to be comparable to that of nearby galaxies, most DLAs tend to
lie at slightly larger radii of the absorbing galaxies for a given $N(\hI)$.

  The extent of neutral gas around intermediate-redshift galaxies may be 
quantified based on the distribution of $M_{AB}(B)$ versus $\rho$ of known 
galaxy-DLA pairs.  The right panel of Figure 1 presents the $M_{AB}(B)$ versus
$\rho$ distribution for 14 galaxy-DLA pairs.  It shows that the data points are
enclosed in an envelope that stretches to larger $\rho$ at brighter 
$M_{AB}(B)$, indicating a finite size of the underlying \hI\ disks.  Chen \& 
Lanzetta (2003) adopted a power-law model to characterize this envelope
\begin{equation}
\frac{R}{R_*} = \left( \frac{L_B}{L_{B_*}} \right)^{\beta}
\end{equation}
and found $R_*=24-30\ h^{-1}\ {\rm kpc}$ and $\beta=0.26-0.29$ for $M_{{\rm 
AB}_*}(B) - 5\log h = -19.6$ (Ellis \etal\ 1996).

  In addition, rotation curve measurements along the major axis of the optical
disk of the galaxies toward PKS0439$-$433 ($z_{\rm DLA}=0.101$), Q0809$+$483 
($z_{\rm DLA}=0.437$), and B2\,0827$+$243 ($z_{\rm DLA}=0.525$) are presented 
in Figure 2.  Note that the observed inner slope of these rotation curves is 
expected to be shallower than the intrinsic shape because of a 
``beam-smearing'' effect caused by seeings that has not been removed from the
data.  Nevertheless, the observations allow us to compare the relative motions
between the ionized gas in the inner ISM and the neutral gaseous clouds at 
large galactocentric distances, and to obtain a robust measurement of the 
rotation speed of galaxy disks at intermediate redshifts.  In all three cases, 
we observe consistent rotation motion between the optical disks and the neutral
gaseous clouds.  These results confirm that galaxies with disk-like stellar
morphologies at intermediate redshifts possess large rotating \hI\ disk out to
$24-30\ h^{-1}$ kpc, a factor of $\approx\sqrt{2}$ larger than what is observed
in local disk galaxies (Rosenberg \& Schneider 2002).

\begin{figure}
\plotfiddle{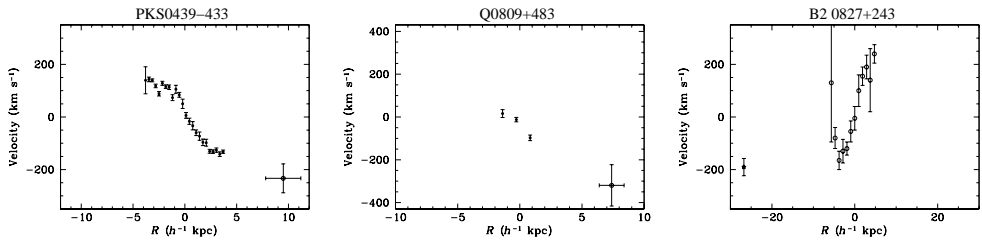}{1.1in}{0}{55}{55}{-190.0}{-10.0}
\caption[]{Rotation velocity measurements of three DLA galaxies versus 
galactocentric radius $R$ along the disk, in comparison to the velocity 
difference between the galaxies and the DLA deprojected to the optical disks
at large $R$.  The velocity measurements presented in the plot has been 
corrected for the inclination of the optical disk.  Measurements for the galaxy
toward Q0827$+$243 were digitized from Steidel \etal\ (2002).}
\end{figure}

\section{Photometric Properties of the DLA Galaxies}

  A comparison of the luminosity distribution of the DLA galaxies and models
derived from adopting the scaling relation of equation (1) and a known galaxy
luminosity function allows us to determine how the neutral gas cross section is
distributed among galaxies of different intrinsic luminosity.  The results are
presented in Figure 3 for 12 confirmed DLA galaxies selected from random lines
of sight (excluding the two systems toward TON\,1480 and HS1543$+$5921; see 
\S\ 2).  Figure 3 shows that the observed luminosity distribution of the DLA 
galaxies agrees well with what is expected from the field galaxy population, if
all field galaxies possess an extended \hI\ envelope described by equation (1).
In addition, it shows that $\approx 30$\% of the DLAs originate in dwarf 
galaxies of $L_B\le 0.25\ L_{B_*}$, $\approx 50$\% of the DLAs originate in
sub-$L*$ galaxies of $0.25\ L_{B_*} \le L_B \le L_{B_*}$, and $\approx 20$\% of
the DLAs originate in super-$L*$ galaxies of $L_B \ge L_{B_*}$.  Including 
candidate DLA galaxies listed in Table 2 does not alter the luminosity 
distribution.

\begin{figure}
\plotfiddle{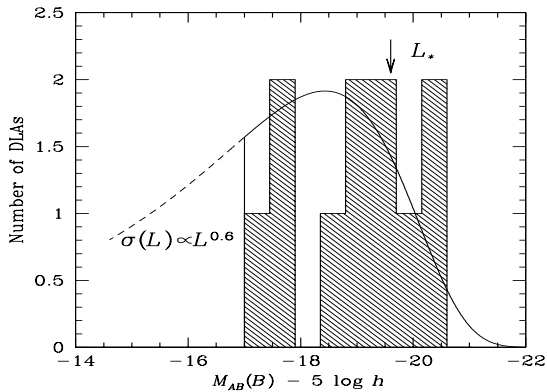}{2in}{0}{40}{30}{-125.0}{-60.0}
\caption[]{Luminosity distribution of DLA galaxies (the shaded histogram), in 
comparison to predictions (the solid curve) from adopting a Schechter 
luminosity fuction, which is characterized by $M_{{AB}_*}(B)=-19.6$ and 
$\alpha=-1.4$, and the best-fit scaling relation.  The model has been 
normalized to match the total number of DLA galaxies observed in the 
homogeneous sample.  The dashed curve indicates the expected number of DLAs 
produced by these fainter galaxies if their neutral gas cross section were to 
be characterized by the same scaling relation of the more luminous ones.}
\end{figure}

  High spatial resolution images of low-redshift DLA galaxies available in the
literature already display a wide range of morphological types among galaxies
that are associated with DLAs (Le Brun \etal\ 1997; Turnshek \etal\ 2001; Chen 
\& Lanzetta 2003).  Figure 4 presents individual images of five additional DLAs
for which moderate resolution spectra of the absorbing galaxies are available 
(see \S\ 5).  The summary in Table 1 indicates that of all the 14 confirmed DLA
galaxies, 43\% are disk dominated, 22\% are bulge dominated, 14\% are 
irregular, and 21\% are in galaxy groups, confirming that DLAs originate in
galaxies of different morphologies and are associated with a variety of galaxy 
environments.

\begin{figure}
\plotfiddle{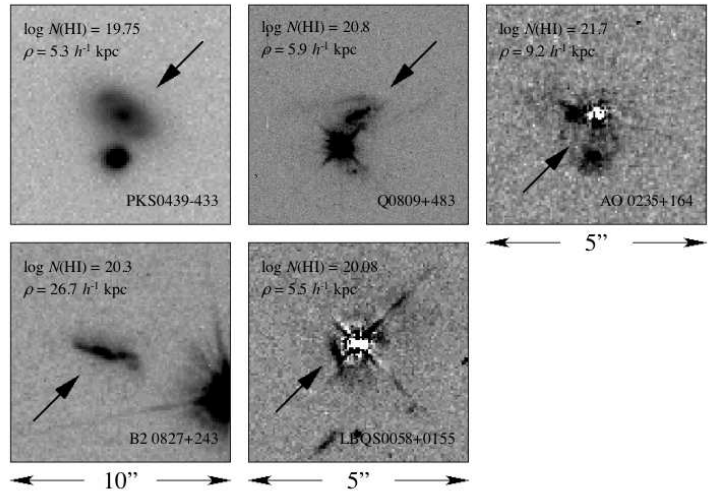}{2.45in}{0}{45}{45}{-125.0}{-90.0}
\caption[]{Direct images of five DLA galaxies at $z<0.65$, showing a range of
morphologies.  The images are 10 arcsec on a side for the fields toward
PKS0439$-$433 and B2\,0827$+$243 and 5 arcsec for the rest.  Field orientation
is arbitrary.  The light from the background QSOs toward AO\,0235$+$164 and 
LBQS0058$+$019 have been subtracted to bring out the faint features of the 
absorbing galaxies (Chen \etal\ 2004).}
\end{figure}

\section{Spectral Properties of the DLA Galaxies}

  Figure 5 shows optical spectra of six DLA galaxies, four of which (toward 
PKS0439$-$433, Q0738$+$313, AO\,0235$+$164, and B2\,0827$+$243) are previously 
known, and two (toward Q0809$+$483 and LBQS0058$+$0155) are confirmed in a 
recent spectroscopic study (Chen, Kennicutt, \& Rauch 2004).  The spectra 
exhibit a range of properties in the ISM, from post-starburst, to normal disks,
and to starburst systems, again supporting the notion that DLA galaxies are 
drawn from the typical field population.

\begin{figure}
\plotfiddle{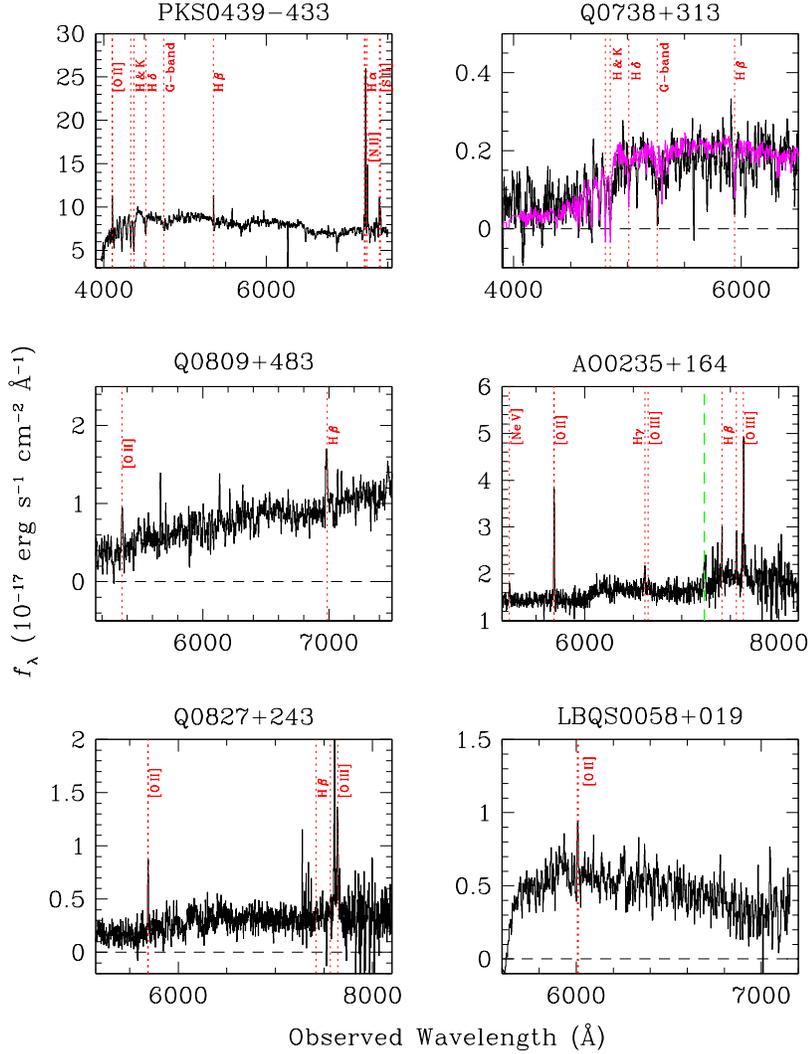}{4.7in}{0}{55}{55}{-165.0}{-25.0}
\caption[]{Moderate-resoultion spectra of six DLA galaxies toward PKS0439$-$433
($z_{\rm DLA}=0.101$), Q0738$+$313 ($z_{\rm DLA}=0.2212$), Q0809$+$483 
($z_{\rm DLA}=0.437$), AO\,0235$+$164 ($z_{\rm DLA}=0.524$), B2\,0827$+$243 
($z_{\rm DLA}=0.525$), and LBQS0058$+$019 ($z_{\rm DLA}=0.525$).  Strong
emission-line features of the galaxies are marked with red dotted lines.  
Contaminating [O\,II] emission feature from the background QSO AO0235$+$164 
($z_{\rm QSO} = 0.94$) is marked with green dashed line.  All but two galaxies 
show spectral features of a typical disk galaxy.  The spectrum of the DLA 
galaxy toward Q0738$+$313 exhibits a pronounced 4000-\AA\ flux discontinuity 
and a strong Ca\,II doublet, in good agreement with a 2.6-Gyr old 
post-starburst model spectrum (the thin magenta curve) generated by the Bruzual
\& Charlot stellar population synthesis code (2003).  The DLA galaxy toward 
AO\,0235$+$164 displays mixed spectral features of both an active nucleus and 
starburst system, as expected from the complex morphology shown in Figure 4.}
\end{figure}

  Strong-line oxygen abundances for three DLA galaxies based on the $R_{23}$ 
calibrator are shown in Figure 6 at galactocentric radius $R\sim 0$.  Although
the $R_{23}$ index may systematically overestimate the oxygen abundance by 
0.2--0.5 dex (Kennicutt \etal\ 2003), the photometric and spectral properties
of these DLA galaxies are consistent with the luminosity and $R_{23}$-based 
metallicity relation of field galaxies (Kobulnicky \& Zaritsky 1999).  
Comparisons with the metallicities of the DLAs at large $R$ suggest that these 
three DLA systems originate in the relatively unevolved outskirts of galactic 
disks.   This in turn supports the scenario in which the low metal abundances 
observed in DLA systems result from a gas cross-section selection bias, which 
favors large impact parameters.

\begin{figure}
\plotfiddle{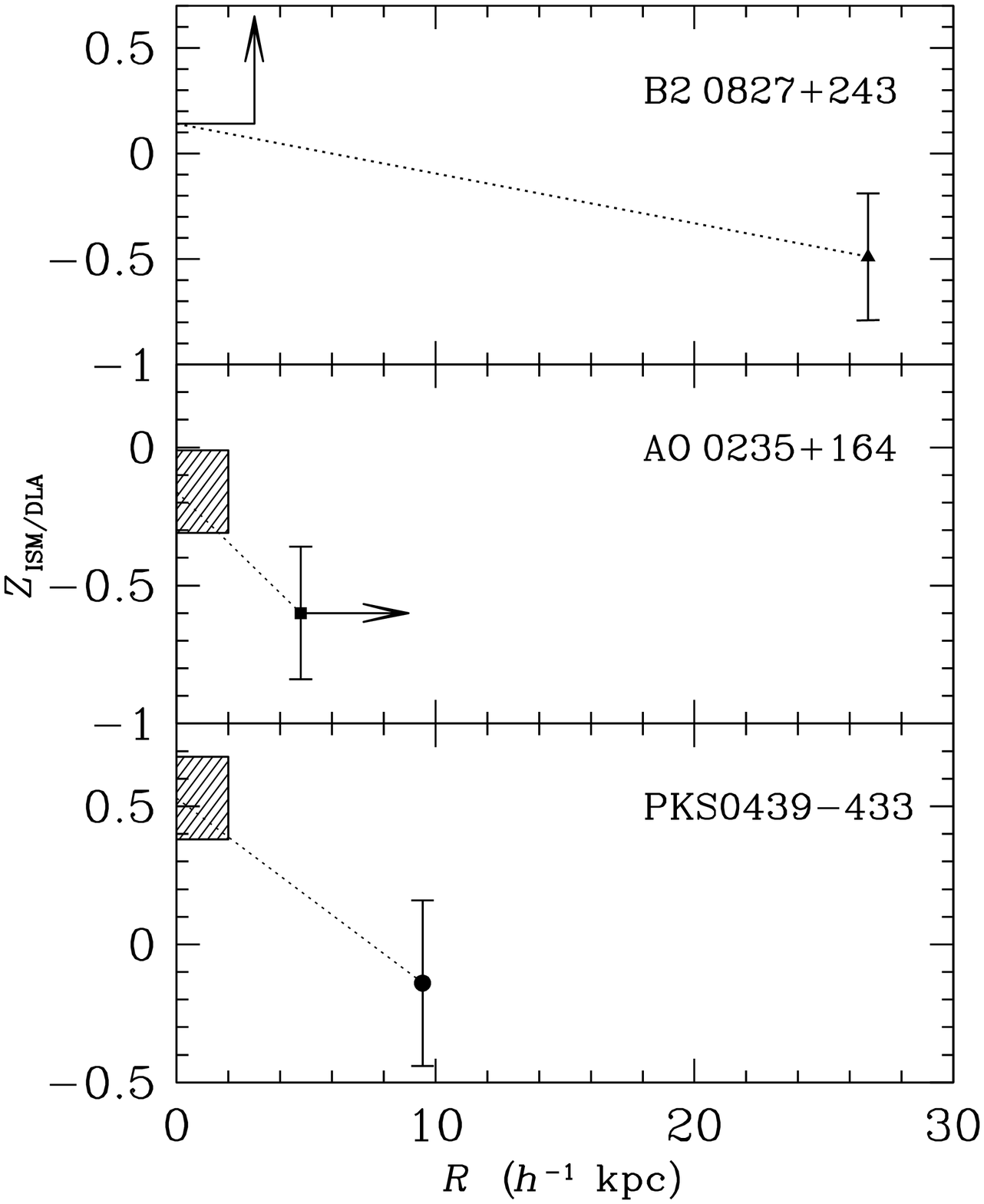}{2.4in}{0}{45}{30}{-140.0}{-25.0}
\caption[]{Abundance decrement observed from the inner ISM of the absorbing
galaxies to the neutral gas at large galactocentric radii for three DLA 
systems.  Shaded box indicate the mean oxygen abundance averaged over the
inner stellar disk, including uncertainties.  The $R_{23}$ calibrator may 
systematically overestimate the oxygen abundance by 0.2--0.5 dex, reducing the
decrement accordingly.  Metallicities of the absorbers are estimated based on 
the observed Fe abundance, corrected for dust depletion (Chen \etal\ 2004).  
All abundace measurements are normalized to their respective solar values 
according to $Z\equiv\log({\rm X}/{\rm H})-\log({\rm X}/{\rm H})_\odot$.}
\end{figure}

\acknowledgements{I would like to thank Rob Kennicutt and Michael Rauch for 
allowing me to discussion some of the results in advance of publication.  This 
research was supported in part by NASA through a Hubble Fellowship grant
HF-01147.01A from the Space Telescope Science Institute, which is operated by
the Association of Universities for Research in Astronomy, Incorporated, under
NASA contract NAS5-26555.}

\end{document}